\documentclass[prc,aps,twocolumn,preprintnumbers,amsmath,amssymb,floatfix,superscriptaddress]{revtex4}
\usepackage[usenames]{color}
\usepackage{amsmath}
\usepackage{graphicx}
\usepackage{floatflt,epsfig}
\usepackage{lscape}
\newcommand\be{\begin{equation}}
\newcommand\ee{\end{equation}}
\newcommand\bea{\begin{eqnarray}}
\newcommand\eea{\end{eqnarray}}

\begin{document}
\title{Population of dipole states via isoscalar probes:
the splitting of pygmy dipole resonance in $^{124}$Sn}
\author{E. G. Lanza}
\email{lanza@ct.infn.it}
\affiliation{INFN Sezione di Catania, Catania, Italy}
\affiliation{Dipartimento di Fisica e Astronomia, Universit\'a di Catania, Italy}

\author{A. Vitturi}
\affiliation{Dipartimento di Fisica e Astronomia, Universit\'a di Padova, Italy}
\affiliation{INFN Sezione di Padova, Padova, Italy}

\author{E. Litvinova}
\affiliation {Department of Physics, Western Michigan University,
Kalamazoo, MI 49008-5252, USA}
\affiliation{National Superconducting
Cyclotron Laboratory, Michigan State University, East Lansing, MI
48824-1321, USA}

\author{D. Savran}
\affiliation{ExtreMe Matter Institute EMMI and Research Division, GSI Helmholtzzentrum f\"ur Schwerionenforschung GmbH, Darmstadt, Germany}
\affiliation{Frankfurt Institute for Advanced Studies, Frankfurt am Main, Germany}

\begin{abstract}

Inelastic $\alpha$-scattering excitation cross sections are calculated
for electric dipole excitations in $^{124}$Sn based on the transition
densities obtained from the relativistic quasiparticle time-blocking
approximation (RQTBA) in the framework of a semiclassical
model. The calculation provides the missing link to directly compare
the results from the microscopic RQTBA calculations to recent
experimental data measured via the $(\alpha ,\alpha '\gamma)$
reaction, which show a structural splitting of the low-lying E1
strength often denoted as pygmy dipole resonance (PDR). The
experimentally observed splitting is reproduced by the cross section
calculations, which allows to draw conclusion on the structure of the PDR.

\end{abstract}

\maketitle


The study of nuclei with neutron excess have been pursued in the last
years with focus to the properties of collective states. Special
attention has been devoted to the electric dipole strength at low
excitation energy, the so called Pygmy Dipole Resonance (PDR), both in
theory \cite{paar07} as well as experimentally \cite{savr13}. While
the PDR represents an interesting new nuclear phenomenon itself, the
connection of the (low-lying) E1 strength to the neutron skin of
atomic nuclei
\cite{piek06,klim07,tson08,paar09,carb10,piek11,vret12,naka13,repk13}
and to isovector parameters in the equation of state of nuclear matter
\cite{rein10,tami11,piek11,piek12} as well as its possible importance
for reaction rates in astrophysical scenarios \cite{gori98,litv09b}
has further increased the interest in the PDR region. Different
microscopic models predict the presence of the low-lying PDR states
below the well-know (isovector) electric giant dipole resonance
(IVGDR), see \cite{paar07,savr13} and references therein. Most
calculations show similar results for the transition densities of the
low-lying E1 strength, which often is used to ``define'' the PDR
states: The neutron and proton transition densities are in phase
inside the nucleus and at the surface only the neutron part
considerably contributes. The investigation of the PDR by means of
isovector (IV) as well as isoscalar (IS) probes provides information
on this structure of the involved transition densities, which show a
strong mixing of isospin character. While for isovector probes
(i.e. photons) data exist for various nuclei obtained mainly in
nuclear resonance fluorescence (NRF) and Coulomb excitation (for a
recent overview see \cite{savr13}), only recently data on the PDR
using an isoscalar probe became available
\cite{savr06,endr09,endr10,endr12,dery13}. Applying the ($\alpha,
\alpha' \gamma$) coincidence method in combination with
high-resolution $\gamma$-ray spectroscopy \cite{savr06b} allowed to
study the PDR for several semi magic stable nuclei in
$\alpha$-scattering experiments. The comparison to results from NRF
revealed a surprising splitting of the PDR strength below the particle
threshold into two groups \cite{savr06,endr10}: the lower lying group
of states is excited by both isoscalar and isovector probes while the
states at higher energy are excited by photons only. A first
qualitative comparison to calculations within the relativistic
quasiparticle time-blocking approximation (RQTBA) as well as
quasiparticle phonon model (QPM) has shown good agreement to the
experimental observation \cite{endr10}: While the low-lying part of
the E1 strength shows the described pattern of transition densities,
which lead to an enhancement in the isoscalar E1 response, the higher
lying states are of transition character towards the GDR and, thus,
are suppressed in the isoscalar channel. Meanwhile, similar IS/IV
energy behaviour of the low-lying E1 strength has been reported
experimentally in inelastic scattering of $^{17}$O off $^{208}$Pb at
low bombarding energies \cite{brac13} as well as in other microscopic
model calculations for different nuclei, see
e.g. \cite{vret12,roca12}. However, the comparison to the experimental
data lacks on the isoscalar part, since the calculated IS B(E1)
strength cannot be directly compared to the measured
$\alpha$-scattering cross sections. The comparison thus remained on
the ``qualitative level''. In this manuscript we present calculation
of $\alpha$-scattering cross sections, based to the microscopic
transition densities obtained in the RQTBA, within the framework of a
semiclassical model to overcome this drawback in the comparison of
experiment and calculation.

The relation between the inelastic cross section and the reduced
transition probabilities B(E1) is clear for the Coulomb excitation or
NRF (they are proportional) while it is not so evident in the relation
between the isoscalar response and the inelastic excitation cross
section due to an isoscalar probe. In this case it is better to
calculate explicitly the inelastic cross section due to the nuclear
interaction in order to establish its connection with the
isoscalar transition probability. Recently, calculation along this
line have been performed \cite{vitt10,vitt11,lanz10,lanz11} within a
semiclassical model. Here we present the result for the system
$\alpha$ + $^{124}$Sn at $E_\alpha$=136 MeV which have been
experimentally studied in refs.~\cite{endr10,endr12}.


The calculation of the inelastic cross sections are performed within a
semiclassical model. The basic assumption is that the colliding nuclei
move on a classical trajectory determined by the real part of the
optical potential, while the internal degrees of freedom are described
quantum mechanically. This model is known to hold for heavy-ion
grazing collisions. In our specific case we consider the alpha particle
 projectile scattered off by $^{124}$Sn at $E_\alpha$=136 MeV and 
analyze the excitation process of the target. Under these conditions, 
we can write the Hamiltonian as
$H_T = H_T^0 + W (t)$
where $H^0_T$ is the internal hamiltonian of the target and the
external field $W$ describes the excitation of T by the mean field of
the projectile nucleus, whose matrix elements depend on time through
the relative coordinate {\bf R}(t).  By solving the Schr\"odinger
equation in the space spanned by the eigenstates of the internal
hamiltonian $|\Phi_\alpha>$ one can calculate, non perturbatively,
the final population for each of the $|\Phi_\alpha>$ states.
Then the time dependent state, $|\Psi(t)>$, of the target nucleus can
be expressed as
\begin{equation}
|\Psi(t)>  = \sum_\alpha A_\alpha (t) e^{-i E_\alpha t} |\Phi_\alpha>
\end{equation}
where the ground state is also included in the sum as the term
$\alpha=0$. The Schr\"odinger equation can be cast into a set of
linear differential equations for the amplitudes $A_\alpha (t)$,
\begin{equation} \label{adot}
\dot A_\alpha (t) = -i \sum_{\alpha^\prime} e^{i (E_\alpha
    - E_{\alpha^\prime}) t} <\Phi_\alpha|W(t)|\Phi_{\alpha^\prime}>
    A_{\alpha\prime} (t) \,.
\end{equation}
Their solutions are then used to
construct the probability of exciting the internal state
$|\Phi_\alpha>$ as
\begin{equation}
P_\alpha(b) = |A_\alpha (t = + \infty)|^2
\end{equation}
for each impact parameter $b$. Finally, the total cross section is
obtain
\begin{equation}
\sigma_\alpha = 2\pi \int_{0}^{+\infty} P_\alpha (b) T(b) b db \> .
\label{sig}
\end{equation}
by integrating $P_\alpha$ over the whole range of the impact
parameters.  The integral is modulated by the transmission coefficient
$T(b)$ which takes into account processes not explicitly included in
the model space and that take flux away from the elastic channel.  
A standard practice is to construct it by integrating
the imaginary part of the optical potential associated
to the studied reaction. When the imaginary part is not available from
the experimental data we use the simple assumption of taking it as
half of the real part.

\begin{figure}[htbp]
\begin{center}
\hspace{-1.0mm}
\includegraphics[angle=0,
width=1.025\columnwidth]{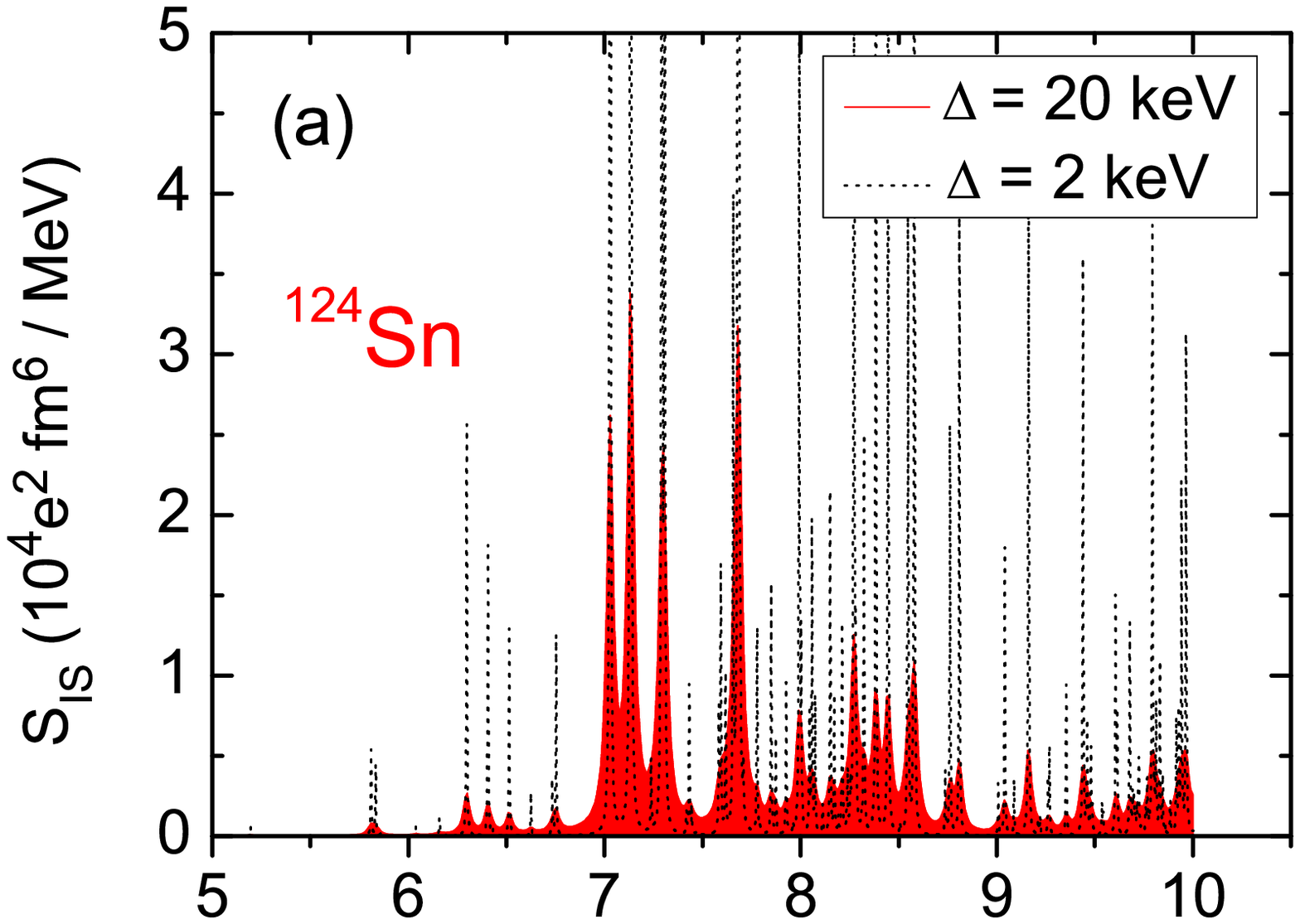}\\
\vspace{-0.5cm}\includegraphics[angle=0,
width=0.9\columnwidth]{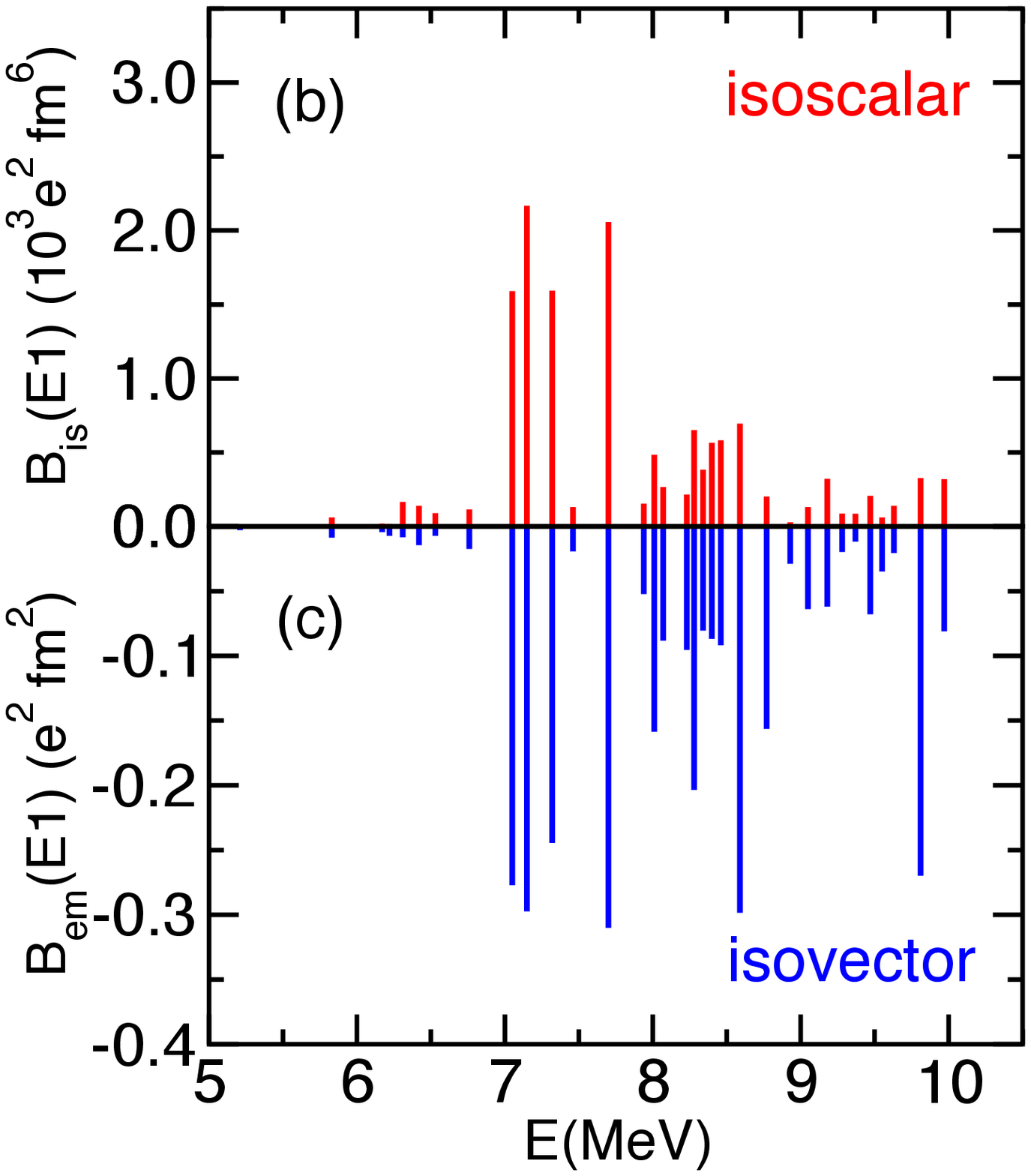} \caption{(Color
online) Bunching procedure for the strength distribution calculated
within RQTBA (a) and the obtained isoscalar (b) and electromagnetic (c)
reduced transition probabilities of the bunched states.}
\label{stre}
\end{center}
\end{figure}

The internal structure of the nucleus $^{124}$Sn is provided by the
relativistic quasiparticle time blocking approximation developed in
refs. \cite{litv08,litv09} and based on the covariant density
functional with NL3 parametrization \cite[]{lala97}.  This approach
is a self-consistent extension of the relativistic quasiparticle
random phase approximation \cite{paar03} accounting for the
quasiparticle-vibration coupling.
The details of the calculations for $^{124}$Sn are given in ref.
\cite{endr12}.  The RQTBA calculation scheme has been approved in
various applications and justified by the full self-consistency and
renormalization technique. The phonon space is truncated by the
angular momentum of the phonons at $J^{\pi} = 6^+$ and by their
frequencies at 15 MeV. The two-quasiparticle space is sufficiently
large to provide decoupling of the translational spurious mode from
the physical ones.
Coupling to vibrations within the RQTBA does not affect this mode
due to the subtraction of the static contribution of the
particle-vibration coupling (PVC) amplitude. This subtraction also
removes double counting of the PVC effects from the residual
interaction, guarantees the stability of the solutions for the
response function and provides fast convergence of the renormalized
PVC amplitude
\cite{tsel13}.  Here we perform a so-called bunching procedure which
is illustrated in Fig. \ref{stre} (a) for the isoscalar dipole
strength distribution. The strength obtained with the smearing
parameter $\Delta$ = 20 keV is transformed into the reduced
transition probabilities via the relation: $B^{\nu} = \pi\Delta
S(\omega_{\nu},\Delta)$, where $S(\omega_{\nu},\Delta)$ is the value
of the strength at the $\nu$-th maximum. The bunched states obtained
in this way accumulate the transition probablilities of all the
states in $\sim$40 keV bins. The distributions of the probabilities
obtained by this procedure are displayed in Fig.  \ref{stre} (b,c)
for the isoscalar and electromagnetic transitions, respectively.
The transition densities of these bunched states are determined as
described in Ref. \cite{litv07b} and will be used in the subsequent
cross section calculations.

The real part of the optical potential, which together with the
Coulomb interaction determines the classical trajectory, is
constructed with the double folding procedure \cite{satc79}, as well as  
the nuclear formfactors by double folding the RQTBA transition densities.
In both cases the nucleon nucleon
interaction has been chosen to be the M3Y-Reid type\cite{bert77}.
Since the excitation is produced by a probe which is essentially
isoscalar and with N=Z, only
the isoscalar part of the nucleon nucleon interaction and the
isoscalar densities and transition densities contribute to the
potential and to the formfactors.  More details are given in
ref.~\cite{lanz11}.

\begin{figure}[htbp]
\begin{center}
\includegraphics[angle=0, width=0.8\columnwidth]{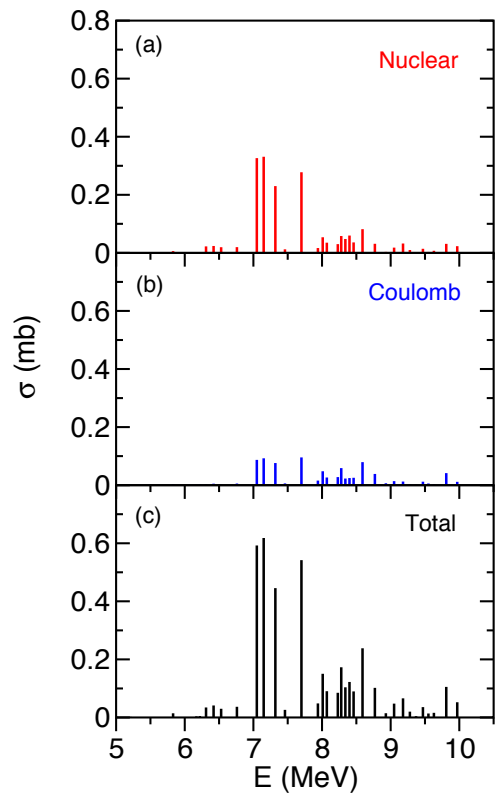}
\caption{(Color online) Inelastic cross sections as function of the excitation
  energy for the systems $\alpha$ + $^{124}$Sn at $E_\alpha$=136 MeV. The
  calculations are performed when only the nuclear (panel (a)) and Coulomb
  (panel(b)) interaction is switched on. The results when both interactions
  are working are in panel (c).
  }
\label{xsec}
\end{center}
\end{figure}

For the given conditions
the important part of the inelastic excitation at E$_{\alpha}$ = 136
MeV is due to the nuclear interaction. Nevertheless, the small
contribution due to the Coulomb interaction is important because of
interference effects. A comparison of pure nuclear and coulomb cross
section as well as total cross section including the interference for
the RQTBA bunched states is given in Fig.~\ref{xsec}. 
The contribution of the Coulomb interaction, panel (b), is small as 
compared to the cross section generated by the nuclear interaction, 
panel (a). When both interaction are switched on an interference effect is
produced, which results in the cross section presented in Fig.
\ref{xsec} (c).  The interference is destructive at small radii and
constructive at large radii and this depends on the different
structure of the nuclear and Coulomb form factor, which in turn are
compelled by the fact that the isoscalar dipole transition density
displays nodes. This is  discussed in a more detailed way in
ref.~\cite{cata97,lanz11}.

The gross features of the strength distributions (see Fig. \ref{stre})
is retrieved in the cross section calculation both for the isoscalar
and isovector cases. Indeed, while we know that this is mathematically
true for the Coulomb case, it is not clear that the same is actually
correct for the nuclear interaction. One can verify it by performing 
a calculation by putting the energies of the
states to zero. This eliminates the contributions due to the dynamic
of the reaction, such as the Q-value effect, and, at least for the
Coulomb case should produce a cross section which is proportional to
the $B_{em}(EL)$ values.

\begin{figure}[htbp]
\begin{center}
\includegraphics[angle=0, width=0.7\columnwidth]{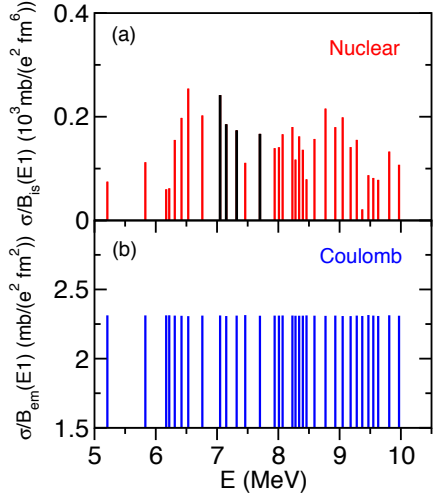}
\caption{(Color online) Ratios between the inelastic cross sections, calculated
by putting to zero the energies of the states, to their corresponding B(EL).
The two panels show the results for the nuclear (panel (a)) and Coulomb
(panel(b)) excitations. The four black lines correspond to the four dipole
states in fig.~\ref{stre} (a) with the bigger response to the isoscalar probe.
    }
\label{ratio}
\end{center}
\end{figure}

Indeed, in panel (b) of Fig.~\ref{ratio} this is shown by plotting for
each state the ratio between the Coulomb inelastic cross section and
the $B_{em}(EL)$ value, calculated by putting to zero the energy of
the corresponding state. As it is expected, the ratio is constant for
all the states. Conversely, in the case of nuclear excitation the
ratios do not have a constant value, as presented in Fig. \ref{ratio}
(a). This corroborate the fact that for the nuclear excitation the
individual cross section depends on the characteristics of the
transition densities and one has to make calculations of the sort
presented here in order do gain useful information on the excitation
process.

\begin{figure}[htbp]
\begin{center}
\includegraphics[angle=0, width=0.9\columnwidth]{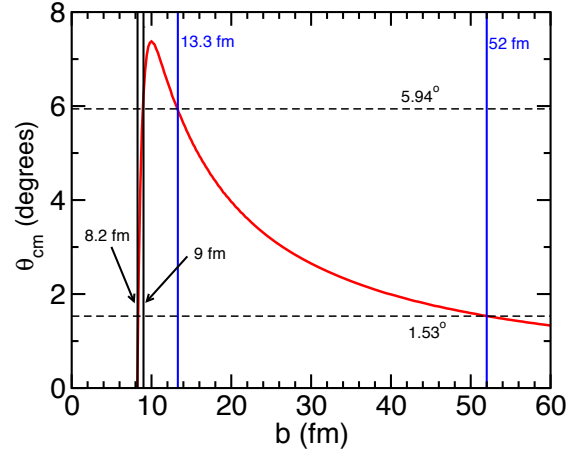}
\caption{(Color online) Deflection function for the system $\alpha$ +
$^{124}$Sn at $E_\alpha$=136 MeV. The horizontal dashed lines delimit
the experimental $\alpha$ scattering angle range. The vertical solid
line determine the corresponding impact parameters ranges.  }
\label{defl}
\end{center}

\end{figure}

\begin{figure}[htbp]
\begin{center}
\includegraphics[angle=0, width=0.8\columnwidth]{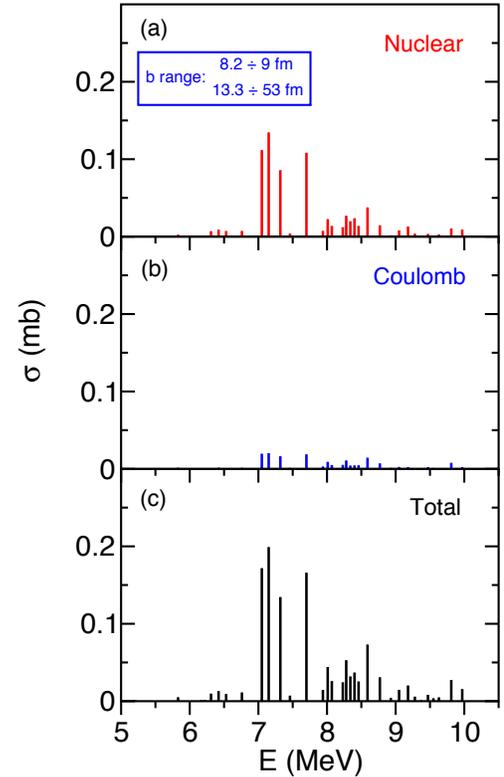}
\caption{(Color online) Same as fig.~\ref{xsec}, in this case only 
the impact parameters intervals whose corresponding
trajectories terminate to the measured scattering angle range are
taken into account.  }
\label{xs-ran}
\end{center}
\end{figure}

So far the presented cross sections are implicitly integrated over the
full solid angle. However, the experimental data of
refs. \cite{endr10,endr12} were taken at the angle range from about
$1.5^\circ$ to $5.5^\circ$ corresponding to about
$1.53^{\circ}$ to $5.94^{\circ}$ in the c.m. system. 
From the deflection function shown
in Fig. \ref{defl} one can deduce the ranges of impact parameters whose
corresponding trajectories will end up to the experimental $\alpha$
scattering angle range. As given in Fig. \ref{defl}, only the impact
parameters between 8.2 fm to 9.0 fm and between 13.3 fm to 52.0 fm
give contributions within the experimental angular range. Therefore,
in order to be consistent with the experiment, the calculations have
been repeated taking into account only the above given impact
parameters ranges. The results are shown in Fig.~\ref{xs-ran}. Apart
from the scale and a small variation, the general features of the three
considered cross sections distributions are maintained and the decrease
of the cross section for energies above 8 MeV is retrieved both for
the nuclear as well as for the total cross sections.


\begin{figure}[htbp]
\begin{center}
\includegraphics[angle=0, width=0.9\columnwidth]{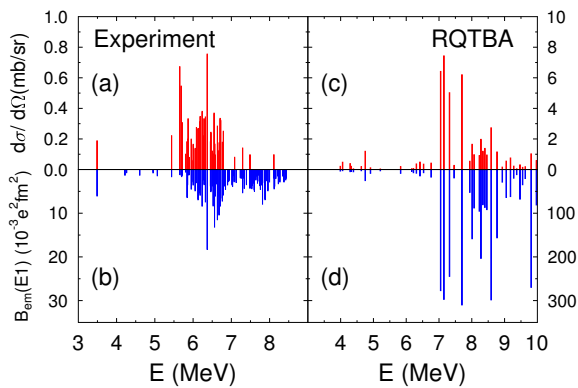}
\caption{(Color online) Comparison of experimental and RQTBA based
cross section for
the $(\alpha ,\alpha '\gamma)$ reaction (upper
row) and B$_{em}$(E1) values (lower row) for $^{124}$Sn. Experimental
values are taken from \cite{endr12,gova98}.  }  \label{comp_dist}
\end{center}
\end{figure}

\begin{figure}[htbp]
\begin{center}
\includegraphics[angle=0, width=0.8\columnwidth]{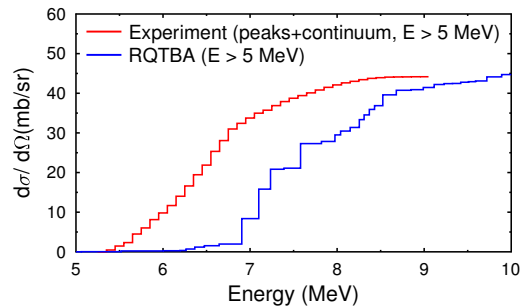} \caption{
\label{comp_run}(Color online) Comparison of experimental and RQTBA
$(\alpha ,\alpha '\gamma)$ cross sections for $^{124}$Sn in terms of
their cumulative sums.  }
\end{center}
\end{figure}

A comparison of the final calculated cross section based on the RQTBA
and the experimental results for $^{124}$Sn is presented in
Fig. \ref{comp_dist}. In order to compare to the differential cross
section measured in the experiment the calculations have been
normalized to the corresponding solid angle. Also given are the
corresponding B$_{em}$(E1) distributions. In both cases the strongest
states are about a factor of ten stronger in the RQTBA compared to the
experiment.
This difference has the following reason. By construction, the
fragmentation in the conventional RQTBA is not sufficient for a
state-by-state description of the dipole strength in this energy
region and coupling to higher configurations have to be included. The
overall ~800 keV shift of the RQTBA strength distributions has the
same origin: all the strength in this energy region consists of
fragments of the RQRPA pygmy mode located at about 9 MeV, and
higher-order correlations are expected to reinforce the fragmentation
to lower energies.  In addition, the bunching procedure described
above results in the strength distributions of ~40 keV bins rather
than of the single states.
However, the similar enhanced factor of about ten that we have for
the cross sections and the B$_{em}$(E1) values for the individual
states shows that the calculations for the cross sections are consistent
and the under-prediction of the fragmentation has no influence on the
gross features of the two distributions.
Integrating the total cross section up to 8.7 MeV results in 43.6
mb/sr for the RQTBA and 44.1(5) mb/sr for the experiment when
including also the contribution of the continuum (see \cite{endr10},
Fig. 3). The cumulative sums of these quantities are displayed in
Fig. \ref{comp_run} showing that the experimental and theoretical
total cross section, aside from the global energy shift, are in
excellent agreement, which further supports the validity of our
calculation on the absolute scale.
However, one has to keep in mind that some details of the calculations
may depend on the uncertainties that are connected with  the use of
the semiclassical model and the associated parameters as well as of
the M3Y-Reid potential.


Looking on the dependence of the strength distribution on excitation
energy a good agreement of data and calculation is observed, i.e. in
both cases the $\alpha$-scattering cross section is strongly reduced
at higher excitation energies compared to the isovector channel. This is
in agreement to the qualitative comparison previously given in Ref.
\cite{endr10}. The proper calculation of the $\alpha$-scattering
excitation cross section, thus, confirms the good agreement between
the experimentally observed splitting of the low-lying E1 strength and
the results of the RQTBA calculation.
However, we are aware of the fact that the NL3 interaction used in
the microscopic structure calculations produce a neutron skin which is
bigger with respect to the experimental data. How this fact can affect
the calculated cross section is not easy to find out and it is out of
the scope of this work. Further work has to be dedicated to this delicate
problem. However, our calculations show, beyond any doubt, that the
same separation is found in the cross section calculations making in
this way the comparison between experiment and theory straightforward.


In summary, we have presented the calculation of $\alpha$-scattering
excitation cross sections for $J^{\pi}=1^{-}$ states based on the
results of microscopic calculation in the RQTBA model using a
semiclassical framework for the reaction.
The aim of this work is to verify that the excitation cross section
due to an isoscalar probe reproduces the structure of the theoretical
isoscalar B(E1), whose comparison with the experimental data was
done in a previous work\cite{endr10} only on a qualitative level.
We have shown that the comparison between experimental and theoretical
cross sections confirms the different behaviour of the population of
the low lying dipole states with different probes.
In the calculations nuclear-Coulomb interference
is included and the resulting cross sections are sensitive to the
character of the transition densities as expected. The calculations
allow for a direct comparison of the RQTBA model to experimental
results obtained in inelastic $\alpha$-scattering on an absolute
scale. The good agreement to the data confirms the accurate description
of the corresponding transition densities of the low-lying E1 strength
in the RQTBA model, which shows for the low-lying group of E1
excitations enhanced neutron contribution on the surface of the
nucleus and an isoscalar behavior in the interior. The combination of
the presented calculations and the experimental data, thus, provides a
first clear identification of this signature often associated to the
pygmy dipole resonance.

\begin{acknowledgments}
  E.L. acknowledges support from the US-NSF grant PHY-1204486 and
  from NSCL; D.S. acknowledges support by the Alliance
  Program of the Helmholtz Association (HA216/EMMI); 
  E.G.L. and A.V. acknowledge support by the italian
  PRIN (grant 2009TWL3MX).
\end{acknowledgments}


\end{document}